\documentclass[12pt]{article}

\def\bs{\begin{subequations}}
\def\es{\end{subequations}}

\catcode`\@=11

\newtoks\@stequation
\def\subequations{\refstepcounter{equation}
  \edef\@savedequation{\the\c@equation}%
  \@stequation=\expandafter{\theequation}
  \edef\@savedtheequation{\the\@stequation}
  \edef\oldtheequation{\theequation}%
  \setcounter{equation}{0}%
  \def\theequation{\oldtheequation\alph{equation}}}

\def\endsubequations{\setcounter{equation}{\@savedequation}%
  \@stequation=\expandafter{\@savedtheequation}%
  \edef\theequation{\the\@stequation}\global\@ignoretrue}

\makeatletter
        \renewcommand{\theequation}{\thesection.\arabic{equation}}%
        \@addtoreset{equation}{section}%
\makeatother

\renewcommand{\thefootnote}{\fnsymbol{footnote}}

\begin{document}
\begin{titlepage}

Revised September 21, 2006   
\begin{center}        \hfill   \\
            \hfill     \\
                                \hfill   \\

\vskip .25in

{\large \bf A Relativistic Quaternionic Wave Equation \\}

\vskip 0.3in

Charles Schwartz\footnote{E-mail: schwartz@physics.berkeley.edu}

\vskip 0.15in

{\em Department of Physics,
     University of California\\
     Berkeley, California 94720}
        
\end{center}

\vskip .3in

\vfill

\begin{abstract}
We study a one-component quaternionic wave equation which is 
relativistically covariant. Bi-linear forms include a conserved 
4-current and an antisymmetric second rank tensor. Waves propagate 
within the light-cone and there is a conserved quantity which looks 
like helicity. The principle of superposition is retained in a 
slightly altered manner. External potentials can be introduced in a 
way that allows for gauge invariance. There are some results for 
scattering theory and for two-particle wavefunctions as well as the 
beginnings of second quantization. However, we are unable to find a 
suitable Lagrangian or an energy-momentum tensor.

\end{abstract}


\vfill

\end{titlepage}

\renewcommand{\thefootnote}{\arabic{footnote}}
\setcounter{footnote}{0}
\renewcommand{\thepage}{\arabic{page}}
\setcounter{page}{1}

\section{Introduction}

Many attempts have been made to consider the extension of the usual 
quantum theory, based upon the field of complex numbers, to quaternions. 
The 1936 paper by Birkhoff and von Neumann \cite{BvN} opened the door 
to this possibility, and the 1995 book by  Adler \cite{Adl} 
covers many aspects that have been studied.

Here is a wave equation that appears to have escaped previous 
recognition.
\begin{equation}
\frac{\partial \psi}{\partial t} i = \textbf{u} \cdot 
\nabla \psi + m \psi j.\label{a}
\end{equation}
The single wavefunction $\psi$ is a function of the spacetime 
coordinates $\textbf{x}, t$.  The usual elementary quaternions $i,j,k$, 
 are defined by 
\begin{equation}
i^{2}=j^{2}=k^{2} = ijk= -1 \label{a1}
\end{equation}
and
\begin{equation}
\textbf{u} \cdot \nabla = i\frac{\partial}{\partial x} 
+ j\frac{\partial}{\partial y} + k\frac{\partial}{\partial 
z}.\label{a2}
\end{equation}
Boldface type is  used to designate a 3-vector.

This combination (\ref{a2}) of elementary quaternions and space derivatives was 
orginated by  Hamilton \cite{Ham} in 1846; it's square is the negative 
of the Laplacian operator.

What one should note about the equation (\ref{a}) is that it employs 
quaternions which multiply the wavefunction on \emph{both} the right 
side and the left side.  This distinction arises from the 
non-commutativity of quaternion algebra and is central to the 
present study.

\section{Other Equations}

There are other quaternionic wave equations one can consider, based upon 
the apparent structural similarities between quaternions and 
relativity. The simplest is
\begin{equation}
\frac{\partial \psi}{\partial t} = \textbf{u} \cdot 
\nabla \psi\label{a3}
\end{equation}
which, when squared, appears as a 4-dimensional Laplace equation, not a wave 
equation.

Going to two-dimensions we construct
\begin{equation}
\left(
\begin{array}{cc}
\frac{\partial}{\partial t} & \textbf{u} \cdot 
\nabla\\
\textbf{u} \cdot \nabla & -\frac{\partial}{\partial t} 
\end{array} \right) \label{a4}
\Psi = m \Psi .
\end{equation}
When this equation is squared, we do get a wave equation, but it 
is for a tachyon.  If one sets $m=0$ in this equation, it can be 
revised to appear as either two copies of the Weyl equation or the 
Maxwell equations (keeping only the imaginary components).

Various authors have shown that the familiar Dirac equation can be 
put into quaternionic form.  This may be done by putting an $i$ to 
the right of $\Psi$ on one side of equation (\ref{a4}) \cite{DLR} or by the use 
of bi-quaternions in a one-component equation \cite{Lan}.  All of those 
representations involve eight real functions - as does the usual 
Dirac equation -  while the basic 
equation of the current study (\ref{a}) involves only four real functions.

There are two other known relativistic equations with four real 
components.  One of these is the Majorana representation of the Dirac 
equation,
\begin{equation}
i\gamma^{\mu}\partial_{\mu}\psi = m \psi
\end{equation}
where all four of the gamma matrices can be made purely imaginary, so 
that one can take all four components of the Dirac wavefunction to be 
real functions of space-time.  Indeed, if we write our quaternionic 
wavefunction as 
\begin{equation}
\psi = \psi_{0} + i\psi_{1} + j \psi_{2} + k \psi_{3},
\end{equation}
and arrange these four real functions as a column vector,
then our equation (\ref{a}) can be put in exactly this Majorana-Dirac 
form. One awkward feature of that formalism is that the usual Dirac 
Lagrangian becomes useless for an action principle, since every 
single term is identically zero.

The other comparison involves the Weyl equation (two complex components), which 
is usually reserved for massless particles. One can map quaternions onto a 
two-dimensional space of complex 
numbers.  The correspondence can be expressed in terms of the familiar 
Pauli matrices $\textbf{u} \rightarrow -i \mathbf{\sigma}$; and the 
wave equation (\ref{a}) can be written in a pseudo-Weyl form as
\begin{equation}
i\frac{\partial \Psi}{\partial t} = -i \mathbf{\sigma} \cdot 
\nabla \Psi + m \sigma_{2} \Psi^{*}.\label{a5}
\end{equation}
In this second example, one also has trouble with the usual Lagrangian 
in that the mass term is identically zero.

Both of these equations, Majorana-Dirac and modified Weyl, are used in 
the building of supersymmetry theories (see, for example, \cite{Wes}), but only 
after one introduces a second set of wavefunctions - with ``dotted'' spinor 
indices.  Thus, they do return to eight real functions, which are, 
furthermore, not simply real functions but elements of a Grassmann  
algebra.

These comparisons leave me without a definitive answer to the question of 
whether the focal 
equation of this paper (\ref{a}) is truly something new in 
theoretical physics.  The work presented here will be to explore this 
quaternionic wave equation on its own terms and see what interesting 
things arise.

\section{Some Properties}

In the usual quantum mechanics there is ``gauge invariance of the first 
kind'': we can replace the complex wavefunction $\psi$ by 
$exp(i\theta)\psi$.  This freedom is also noted by saying that there is a 
ray, not just one vector, in Hilbert space corresponding to each 
physical state. (The reader will note that this paper focuses entirely 
on the wavefunction approach to quantum theory and not the Hilbert space version.) 
For the quaternionic wavefunction we have a larger 
set of freedoms: $ \psi \rightarrow q_{1}\psi q_{2}$, where the two 
numbers $q_{1}, q_{2}$ are quaternions of unit magnitude.  The one on 
the left induces a change of basis in the elementary quaternions 
$\textbf{u}$ seen in the equation (\ref{a}), while the one on the 
right changes the particular choice of $i$ and $j$ acting to the right 
of $\psi$ in that equation.  Thus, instead of the usual $U(1)$ group, we 
appear to have $SU(2)\times SU(2)/Z_{2}$.

A first calculation is to take another time-derivative of equation 
(\ref{a}) and arrive at the second-order wave equation,
\begin{equation}
\frac{\partial^{2}\psi}{\partial t^{2}}  = \nabla ^{2} 
\psi - m^{2}\psi \label{b}
\end{equation}
which is the ordinary Klein-Gordon equation for a relativistic 
particle of mass m. 

Now we look at some bi-linear forms. The first is $\rho = \psi^{*}\psi$, 
where the complex conjugation operator $(^{*})$ changes the sign of 
each imaginary quaternion (and requires the reversal of order in 
multiplication of any expression upon which it operates). 
The second is the vector $\textbf{U} = 
\psi^{*} \textbf{u} \psi $. While $\rho$ is real, $\textbf{U}$ is 
purely imaginary and we can write $\textbf{U} = i \textbf{U}_{1}
+ j \textbf{U}_{2} + k \textbf{U}_{3}$, in terms of three real 
three-vectors.

Making use of the wave equation (\ref{a}), we then calculate
\begin{equation}
\frac{\partial \rho}{\partial t} = \nabla \cdot
 \textbf{U}_{1},\label{b1}
\end{equation}
which is the familiar statement of a conserved current.  We 
shall return to $\textbf{U}_{2}$ and $\textbf{U}_{3}$ shortly. (If 
you ask what singled out $\textbf{U}_{1}$ as the conserved current, 
it is the choice of the imaginary $i$ sitting beside the 
time-derivative in the wave equation (\ref{a}).)

\section{Space-time Symmetries}

Now we look at the behavior of the wave equation (\ref{a}) under 
familiar symmetry transformations.  To achieve rotation of the spatial 
coordinates $\textbf{x}$, we make the transformation
\begin{equation}
\psi \;\longrightarrow \;e^{R}\; \psi, \;\;\; R= \textbf{u} \cdot
 \mathbf{\theta}/2 \label{c}
\end{equation}
where $\mathbf{\theta}$ is the axis and the angle of rotation.

For the Lorentz transformation, we start with the infinitesimal form,
\begin{equation}
\psi \;\longrightarrow \;\psi + B\; \psi\; i, \;\;\; B = \textbf{u} \cdot
 \textbf{v}/2 \label{c1}
 \end{equation}
 where \textbf{v} is the direction and amount of the velocity boost.  
 Note the appearance of the imaginary $i$ acting on the right of 
 $\psi$ in this transformation.  I leave it as an exercise for the 
 reader to show that this transformation of $\psi$ does indeed induce 
 the familiar Lorentz transformation of the spacetime coordinates in 
 the wave equation (\ref{a}).
 
 One can now readily show that the components of the conserved 
 current ($\rho$ and $\textbf{U}_{1}$) tansform as a Lorentz 
 4-vector.  With a bit more work, one can also see that the other two 
 vectors $\textbf{U}_{2}$ and $\textbf{U}_{3}$ transform as the 
 components of an antisymmetric second rank tensor in 4-dimensions 
 (also called a six-vector).
 
 A useful notation for operators that may multiply quaternionic 
 functions on the right or on the left is the following.\footnote{A 
 similar notation was introduced by the authors of reference \cite{DLR}.}
 \begin{equation}
 (a \| b) \psi = a\;\psi\;b, \;\;\; (a \| b)\;(c \| d) = (ac \| db),\label{c2}
 \end{equation}
 which allows us to write the finite Lorentz transformation 
 operator as $e^{(B \| i)}$.

The generators of the Lorentz Group may be constructed as:
\begin{equation}
\textbf{J} = \textbf{x} \times \nabla -
 \frac{1}{2} \textbf{u} \label{c3}, \;\;\;\;\;
\textbf{K} = \textbf{x} \frac{\partial} {\partial t}+t 
\nabla
- \frac{1}{2} \textbf{u} \| i .
\end{equation}
One can extend this to the full Poincare group by adding the 
displacement operators: $\partial_{\mu} = (\partial_{t},\nabla)$.

In Appendix A is a more extensive study of various tensors that can be 
built from solutions of the wave equation.

\section{More Bilinear Forms}

Start by defining the derivative operator 
which acts in both directions, 
$d_{\mu} = (d_{0},\textbf{d}) =  \frac{1}{2}(\overrightarrow{\partial}_{\mu}  -
\overleftarrow{\partial}_{\mu})$. This is  a covariant  
four-vector, but let us  now see how things behave when we combine 
it with the Lorentz transformation of the wave function:
\begin{equation}
D_{\mu} \equiv  \psi^{*}d_{\mu}\psi
 \rightarrow D_{\mu} + \frac{1}{2}
\left\{i,\psi^{*}d_{\mu}\; \textbf{v} \cdot \textbf{u} \psi 
\right\}.\label{c4}
\end{equation}
The expression $D_{\mu}$ is pure imaginary; and so we can write 
$D_{\mu} =  D_{1,\mu}\;i +  D_{2,\mu}\;j +  D_{3,\mu}\;k$. 
I hope that the use of the subscripts (1,2,3), denoting which 
imaginary component they come from, does not cause confusion with the vector or 
tensor subscripts $\mu$.
 The expression inside the 
anti-commutator brackets (next to i) is real. This leads us to 
conclude that, under the Lorentz transformation of $\psi$:
\begin{eqnarray}
D_{2,\mu}\; and \; D_{3,\mu} \;are \;unchanged \label{c5}\\
D_{1,\mu} \rightarrow D_{1,\mu} + more\;\; complicated\;\; 
stuff.\label{A3}
\end{eqnarray}

This means that under the full Lorentz transformation of both 
coordinates and wavefunction 
 $D_{2}$ and $D_{3}$ behave simply 
as 4-vectors.  The 
quantity $D_{1}$, however,  will be shown in 
Appendix A to be part of a higher rank tensor.

Before proceeding, we note that $D_{\mu = 0}$ can be 
reexpressed by using the wave equation (\ref{a}):
\begin{equation}
D_{\mu=0} = -i \tau  - \frac{j}{2} \nabla \cdot \textbf{U}_{3} + k[m\rho + 
\frac{1}{2} \nabla \cdot \textbf{U}_{2}],\label{c6}
\end{equation}
where $\tau \equiv   \psi^{*}\textbf{u} \cdot \textbf{d} \psi$ 
is a real 3-scalar. Under the Lorentz transformation of the 
wavefunction, we calculate $\tau \rightarrow \tau + \textbf{v}\cdot 
\textbf{D}_{1}$. 

We  have the identity
\begin{equation}
\partial^{\mu} D_{\mu} =  0 ;\label{c6a}
\end{equation}
and we will be interested in the following time derivatives, which 
are derived by using the wave equation (\ref{a}).
 \begin{eqnarray}
&& \frac{\partial }{\partial t} \tau  =  - \nabla \cdot 
\textbf{D}_{1}\label{c7}\\
&& \frac{\partial}{\partial t}D_{\mu}  =  i[ 2m D_{2,\mu}-\nabla \cdot 
(\psi^{*}\textbf{u}d_{\mu}\psi) ] -2mj D_{1,\mu} 
 +[i,\psi^{*}d_{\mu}\textbf{u}\cdot\textbf{d}\psi]\label{c8}\\  
&&\frac{\partial}{\partial t}\textbf{U}  =  i[\nabla \rho + 2 
 \psi^{*}\textbf{u} \times \textbf{d}\psi + 
 2m\textbf{U}_{2}] \nonumber \\
&& + j[-2m \textbf{U}_{1} + 2 \textbf{D}_{3}- \nabla \times \textbf{U}_{3}]
 + k[-2 \textbf{D}_{2} + \nabla \times \textbf{U}_{2}].\label{c9
 }
 \end{eqnarray}
 
 See Appendix A for a more systematic discussion of tensor quantities.

\section{Plane Waves}

One way of representing ``plane-wave'' solutions of the 
wave equation (\ref{a}) is
\begin{equation}
\psi(\textbf{x},t) = exp(\eta \textbf{u} \cdot \hat{p}\; \textbf{p}
\cdot \textbf{x}) \; \phi \;exp((i\eta p + km)t) \label{d}
\end{equation}
where $\eta = \pm 1$.  The set of possible momentum vectors 
$\textbf{p} = \hat{p} p$ should cover only one-half of space, to avoid 
overcounting if solutions.  With this, one can construct the solution 
for the general initial value problem:
\begin{eqnarray}
\psi(\textbf{x},t) =\int d^{3}\textbf{x}^{\prime}\;\;\sum_{\eta}\int_{H} 
\frac{d^{3}p}{(2\pi)^{3}} \; exp(\eta \textbf{u} 
\cdot \hat{p}\; \textbf{p}
\cdot (\textbf{x}-\textbf{x}^{\prime}))\; \nonumber \\ 
\psi(\textbf{x}^{\prime},t^{\prime}=0) \;\;
exp((i\eta p + km)t) \label{d1} 
\end{eqnarray}
where the subscript $H$ reminds us that the integral covers only half of 
momentum space.

With the expansions
\begin{eqnarray}
 exp(\eta \textbf{u} \cdot \hat{p}\; \textbf{p}
\cdot (\textbf{x}-\textbf{x}^{\prime})) = \cos(\textbf{p}
\cdot (\textbf{x}-\textbf{x}^{\prime})) + \eta \textbf{u} 
\cdot \hat{p} \;\sin (\textbf{p}
\cdot (\textbf{x}-\textbf{x}^{\prime})) \label{d2}\\
 exp((i\eta p + km)t) = \cos(\omega t) + 
(i\eta p + km)\; \sin(\omega t)/\omega,\;\;\;\;\;\;\;\;\;\;\label{d3}
\end{eqnarray}
where $\omega = \sqrt{p^{2} + m^{2}}$, we  sum 
over $\eta$ and reduce (\ref{d1}) to the following.
\begin{eqnarray}
\psi(\textbf{x},t) =\int d^{3}\textbf{x}' \int_{H}\;\frac{d^{3}p}{(2\pi)^{3}} \;2[ \cos( \textbf{p}
\cdot (\textbf{x}-\textbf{x}^{\prime}))\; \psi\;(\textbf{x}^{\prime},0) 
\;(\cos(\omega t) \nonumber \\
+ km \sin(\omega t)/\omega) + \textbf{u} 
\cdot \textbf{p} \;\sin (\textbf{p}
\cdot (\textbf{x}-\textbf{x}^{\prime}))\;\psi(\textbf{x}^{\prime},0)\;
i\sin(\omega t)/\omega ]. \label{d4} 
\end{eqnarray}

Here we can recognize that the results of the integrals over 
$\textbf{p}$ (which now may be extended to cover the full momentum space) 
give us functions of the invariant $R^{2} = t^{2} - (\textbf{x} - 
\textbf{x}^{\prime})^{2}$, which vanish outside the light-cone 
($R^{2} > 0$). Thus we do have relativistic causality for this 
quaternionic wave equation; something which we could have expected 
because the solutions satisfy the Klein-Gordon equation.

The Klein-Gordon equation also has the property that positive 
(negative) frequency solutions propagate only to positive (negative) 
frequency solutions. For the quaternionic equation, we have no way to 
talk about this distinction between positive and negative frequencies; 
however we do find a substitute ``selection rule'' for wave 
propagation here. 

First, we note the orthogonality relation,
\begin{equation}
\int \frac{d^{3}x}{(2\pi)^{3}}\;\; exp(-\eta^{ \prime} \textbf{u} \cdot 
\hat{p}^{\prime}\; \textbf{p}^{\prime} \cdot \textbf{x})\;\;
exp(\eta \textbf{u} \cdot 
\hat{p}\; \textbf{p} \cdot \textbf{x})\; = \delta_{\eta, 
\eta^{\prime}}\;\;\delta(\textbf{p} \pm \textbf{p}^{\prime}),\label{d5}
\end{equation}
where I have not required that both sets of momentum variables belong to the 
same half-space.  Next, we use this orthogonality in equation (\ref{d1}), where we represent 
$\psi(\textbf{x}^{\prime},0)$ as any superposition of plane wave 
solutions with exclusively $\eta^{\prime} = +1$ (or exclusively  
-1).  The resulting $\psi(\textbf{x},t)$ will contain only that same  
 value for $\eta$.  It is tempting to call this ``helicity 
 conservation'' in the propagation of these quaternionic waves.
 
 This interpretation is bolstered by the following observations.  The 
 operator $\textbf{u} \cdot \nabla $, acting on a plane wave 
 solution (\ref{d}), has eigenvalue $-\eta p$. Furthermore, one can 
 readily show, from (\ref{a}),  that
 \begin{equation}
 \frac{d}{dt} \int d^{3}x \; \psi^{*} \textbf{u} \cdot 
 \nabla \psi = 0.\label{d6}
 \end{equation}

\section {Superposition} 

In the usual (complex) quantum theory, if we have two solutions to 
the Schrodinger (time-dependent) equation, $\psi_{1}$ and $\psi_{2}$, 
then any linear combinmation $c_{1}\psi_{1}+c_{2}\psi_{2}$ is also a 
solution, for arbitrary complex numbers $c_{1}$ and $c_{2}$. With our 
quaternionic wave equation (\ref{a}), the idea of superposition requires a 
slightly different wording.

Note that the general plane wave solution (\ref{d}) has an arbitrary 
amplitude $\phi$ positioned in the midst of certain quaternionic 
functions of space and time. Given any such solution, we find another 
solution by changing the amplitude: $\phi \rightarrow q\phi q\prime$, where 
$q$ and $q\prime$ are arbitrary quaternionic numbers. Furthermore, if 
we have one solution of (\ref{a}) - $\psi_{1}$ with amplitude 
$\phi_{1}$ - and another solution - $\psi_{2}$ with amplitude 
$\phi_{2}$ - then we also have a solution by simply adding these 
two: $\psi_{1} + \psi_{2}$. This version of the principle of 
superposition is implicit in equation (\ref{d1}).

\section{Adding Potentials}

 The original wave equation (\ref{a}) can be extended by the 
 introduction of external potentials, as follows.
 \begin{equation}
 \frac{\partial \psi}{\partial t} i = \textbf{u} \cdot 
\nabla  \psi + e\varphi\psi - e \textbf{u} \cdot \textbf{A} 
\psi i + m \psi e^{ieW}j \label{e}
\end{equation}
where $\varphi, \textbf{A}, W$ are real functions of spacetime.
The gauge transformation that leaves this equation invariant is:
\begin{eqnarray}
\psi \longrightarrow \psi e^{ie\chi}\label{e1} \\
\varphi \longrightarrow \varphi - \frac{\partial \chi} {\partial t}\label{e2} \\
\textbf{A} \longrightarrow \textbf{A} + \nabla \chi 
\label{e3}\\
W \longrightarrow W - 2\chi.\label{e4}
\end{eqnarray}

One can show that the previously discussed symmetries still hold, 
with $(\varphi,\textbf{A})$ a Lorentz 4-vector and $W$ a scalar. This 
appearance of the 4-vector potentials is (almost) exactly like the usual way of 
introducing 
electromagnetism into quantum theory; however, the explicit appearance 
of a gauge 
quantity $W$ is something different. 

The reflection symmetries of equation (\ref{e}) are:
\begin{eqnarray}
\psi \rightarrow \psi\;j, \;\;\; t, \textbf{A}, W \;change \;
sign\;\;\;(T) \label{e5}\\
\psi \rightarrow \psi\;k, \;\;\; \textbf{x}, \varphi, W \;change \;
sign\;\;\;(CP) \label{e6}\\
\psi \rightarrow \psi\;i, \;\;\; t, \textbf{x}, \textbf{A}, \varphi \;change 
\;sign\;\;\;(TCP).\label{e7}
\end{eqnarray}

The current conservation equation (\ref{b1}) is still true for this extended 
wave eqation (\ref{e}); however, equation (\ref{d6}) must be modified.  
For the situation where $W=0$, we calculate
\begin{equation}
\frac{d\tau}{dt} =\frac{d}{dt}(\psi^{*}\textbf{u}\cdot \textbf{d} \psi) = - \nabla \cdot \textbf{D}_{1} + e\rho \nabla 
\cdot \textbf{A} - 2 e \textbf{A}\cdot (\psi^{*}\textbf{u}\times 
\textbf{d}\psi)\label{e8}
\end{equation}
where we have used the notations from Section 5. 
From this we see that in the case where the only external 
potential is $\varphi$, then the space integral of $\tau$,  which we identified with 
 helicity, is conserved.

\section{More on Plane Waves}

The plane wave solutions to the wave equation (\ref{a}), which we set out in 
Section 6, contain an amplitude $\phi$ which we should study some more.
\begin{equation}
\psi(\textbf{x},t) = exp(\eta\;\textbf{u}\cdot 
\hat{p}\;\textbf{p}\cdot \textbf{x})\;\phi\;exp(\Omega  t)\label{db}
\end{equation}
where $\Omega = \hat{\Omega}\omega = (i\eta p + km)$, 
$\omega=\sqrt{p^{2}+m^{2}}$.

We can ask to evaluate the various bilinear forms discussed earlier in 
the case of this plane wave solution. The easiest are 
\begin{equation}
\rho = \phi^{*}\phi, \;\;\; \tau =-\eta p \rho, \;\;\; D_{0} = \Omega 
\rho \label{db1}
\end{equation}
but to do more we must be able to evaluate 
$\phi^{*}\textbf{u}\cdot\hat{p}\phi$.

I now propose to classify the constants $\phi$ in a particular way.  The set 
$\phi_{\alpha}$ are defined such that they perform a specific 
rotation, as follows:
\begin{equation}
\hat{p}\cdot \textbf{u}\;\phi_{\alpha} = 
\phi_{\alpha}\hat{\Omega}\label{db2}
\end{equation}
which sends one unit imaginary quaternion into another. With this 
type, the solution can be written as
\begin{equation}
\psi = \phi_{\alpha}\;exp(\hat{\Omega}(\omega t + \eta 
\textbf{p}\cdot\textbf{x}))\label{db3}
\end{equation}
which looks  like the sort of plane waves we are used to. It should 
be noted that this definition of $\phi_{\alpha}$ is not unique but 
leaves us with a $U(1)$ class of equivalent amplitudes,
\begin{equation}
\phi_{\alpha} \rightarrow \phi_{\alpha} \; exp(\theta 
\hat{\Omega}),\label{db4}
\end{equation}
just as in ordinary (complex) quantum theory.

With this $\alpha$ type of amplitude, we can now evaluate the plane 
wave values for the following bilinears:
\begin{equation}
\textbf{U}\cdot \hat{p} = i\rho \eta \frac{p}{\omega} + k\rho 
\frac{m}{\omega}, \;\;\;\;\; \textbf{D} = \eta\textbf{p}\;
(\textbf{U}\cdot \hat{p}) .\label{db5}
\end{equation}
Components of the vector $\textbf{U}$ which are orthogonal to 
$\textbf{p}$ will oscillate rapidly in space; thus any space average 
of them will be vanishing small.

Two other categories for the amplitudes $\phi$, called $\beta$ 
and $\gamma$, can be defined as
\begin{eqnarray}
\textbf{u}\cdot \hat{p}\;\phi_{\beta} = \phi_{\beta}\;j\label{db6}\\
\textbf{u}\cdot \hat{p}\;\phi_{\gamma} = 
\phi_{\gamma}\;j\hat{\Omega}.\label{db7}
\end{eqnarray}
Note that the three numbers $\hat{\Omega},\;j\;, j\hat{\Omega}$ are 
mutually anticommuting quaternions. If we calculate any of the 
bilinears involving $\textbf{u}\cdot \hat{p}$ with either the $\beta$ or $\gamma$ type 
of amplitude, the result will be rapidly oscillating in time; thus 
any time average will be vanishing small.

If we stay with the $\alpha$ type amplitudes, we get the following 
values, in the plane wave states,  for various 4-vectors that are 
defined in Section 5 or Appendix A.
\begin{eqnarray}
&&j_{\mu} = \rho(1, \eta \textbf{p}/\omega) \label{db8}\\
&&V_{\mu} = \rho(\eta p, \omega \hat{p}) \label{db9}\\
&&D_{2,\mu} = 0 \label{db10}\\
&&D_{3,\mu} = \rho m (1, \eta \textbf{p}/\omega)\label{db11}
\end{eqnarray}
The 4-vectors $j$ and $D_{3}$ look like what we would expect for the 
usual energy-momentum. The 4-vector $V$, however, is 
spacelike, not timelike; it is similar to the spin vector $s_{\mu} 
= \epsilon_{\mu,\nu,\kappa,\lambda}P^{\nu}S^{\kappa,\lambda}$ in the 
usual theories, where $s_{0}$ is the helicity.

The plane wave solutions are characterized by a parameter 
$\textbf{p}$ which we sometimes call ``momentum''. This is merely a 
linguistic 
habit carried over from conventional quantum theory (following 
deBroglie's rule that momentum equals Planck's constant divided by 
wavelength) and should not be confused with the physical quantity 
called  
momentum until and unless that connection is established.

\section{Scattering}

Lets start with the wave equation plus a source
\begin{equation}
\frac{\partial}{\partial t} \psi i = \textbf{u} \cdot 
\nabla \psi + m \psi j + s(\textbf{x},t)i,\label{eb}
\end{equation}
and write the retarded solution as
\begin{equation}
\psi(\textbf{x},t)=\int_{-\infty}^{t}dt' \int d^{3}x' 
\int_{H}
\frac{d^{3}p}{(2\pi)^{3}}\sum_{\eta} exp(\eta \textbf{u}\cdot \hat{p}\textbf{p} 
\cdot(\textbf{x} - \textbf{x}')) s(\textbf{x}',t')
exp(\Omega(t-t'))\label{eb1}.
\end{equation}

For the general scattering problem, we replace the source $s$ 
with $V\psi$ and add in the initial (free particle) solution 
$\psi_{0}(\textbf{x},t)$. If the 
interaction $V$ is independent of time, then we have an integral 
equation,
\begin{eqnarray}
\psi(\textbf{x},t)=\psi_{0}(\textbf{x},t)+ \int_{-\infty}^{t}dt' \int d^{3}x' 
\int_{H}
\frac{d^{3}p}{(2\pi)^{3}}\sum_{\eta} exp(\eta \textbf{u}\cdot \hat{p}\textbf{p} 
\cdot(\textbf{x} - \textbf{x}'))\nonumber\\ V(\textbf{x}')\psi(\textbf{x}',t')
exp(\Omega(t-t')).\;\;\;\;\;\label{eb2}
\end{eqnarray}

Now we make the ``Born approximation'' that $\psi=\psi_{0}$ under 
the integral and let the time $t$ go to $+\infty$. Then, we find that 
the integral over $t'$ gives us a $\delta(\omega - \omega_{0})$, 
which is usually read as conservation of energy.  This result appears 
to be generally true, not just in the first Born approximation.  One 
can now project this solution onto any plane wave solution and 
achieve the quaternionic version of the S-matrix.

In the special case when the scattering potential $V$ comes from the 
term $\varphi$ in the extended wave equation (\ref{e}), we also find - as a 
result of the integral over $t'$ - that we have the selection 
rule $\eta = \eta_{0}$.  This is consistent with the result noted 
after equation (\ref{e8}).

\section{Some Other Solutions}

We can write solutions for the extended wave equation (\ref{e}) in some special 
cases.

One may ask whether there is a central potential, $\varphi(r)$, which 
leads to bound states. The easiest way to explore this is through ``reverse 
engineering'': write down a plausible wavefunction and see what potential fits 
the wave equation. The form
\begin{equation}
\psi = [f(r) + \textbf{u}\cdot \hat{r}\;g(r)]\phi(t)\label{f1}
\end{equation}
leads to the requirements
\begin{equation}
r^{4}\frac{d}{dr} f^{2} = - \frac{d}{dr} (r^{4}g^{2}), 
\;\;\;\;\;e\varphi = -f'/g,\;\;\;\;\;\phi(t) =\phi_{0}e^{kmt}.\label{f2}
\end{equation}
If we try the asymptotic ($r\rightarrow \infty$) behavior, $g \rightarrow
\alpha r^{-\beta}$, we find a similar behavior for $f$, provided 
that $0 < \beta < 2$.  The wavefunction is then normalizable for 
$\beta > 1.5$; and the potential is $e\varphi(r) = 
\sqrt{\beta(2-\beta)}/r$ at large $r$. Looking instead at $r 
\rightarrow 0$, one can do the same analysis and require $\beta < 
1.5$; this suggests we are dealing with something like a shielded 
Coulomb potential.

There are familiar procedures for taking the non-relativistic limit 
of the Klein Gordon or Dirac equations.  Here is the best I could do 
with the present relativistic equation. First, write $\psi = \psi_{nr} 
exp(k\omega t)$, where $\omega = \sqrt{m^{2}+p^{2}} \approx 
m - \nabla^{2}/2m$. Next, 
multiply the equation from the right with $i exp(-k\omega t)$. 
Finally, drop all terms that oscillate rapidly in time, as $exp(\pm 2 
k \omega t)$.  The resulting version of the full extended equation 
(\ref{e}) 
is 
\begin{eqnarray}
&&\frac{\partial\psi_{nr}}{\partial t} k \approx H_{nr}\psi_{nr}\label{f3}\\
&&H_{nr} = -\nabla^{2}/2m +  m (1-cos(eW)) -e\textbf{u}\cdot\textbf{A}||k 
\label{f4}
\end{eqnarray}
which looks like an ordinary Schrodinger equation except that the 
single imaginary is called k instead of i; and there is also the 
unfamiliar term with $\textbf{A}$. What looks like an effective 
potential energy term (coming from the gauge quantity W) is positive, 
thus incapable of producing bound states, although it might 
conceivably yield metastable states through the oscillation of the 
cosine function.

\section{Two Particle Equation}

Previous studies of quaternionic quantum theory have gotten into 
trouble when they try to write wavefunctions for multi-particle 
systems. In the ordinary (complex) theory, one simply makes a direct 
product of one-particle wavefunctions; and because all the numbers 
there commute, one can manipulate such a product to achieve various 
sensible results. In the quaternion case that approach leads to a 
horrid mess. (See, for example, Adler's book, Chapter 9.)

The present work suggests a somewhat different approach.  Consider 
this construction with plane waves: 
\begin{eqnarray}
\psi(1,2) = exp(\eta_{1} \textbf{u}\cdot \hat{p_{1}}\textbf{p}_{1} 
\cdot\textbf{x}_{1})\; \psi(2)\; exp(\Omega_{1}t_{1}) \label{g}\\
\psi(2) = exp(\eta_{2} \textbf{u}\cdot \hat{p_{2}}\textbf{p}_{2} 
\cdot\textbf{x}_{2}) \;\phi\; exp(\Omega_{2}t_{2})\label{g1}
\end{eqnarray}
which might be described as a ``nested'' product. The symbol $\phi$ 
here is a quaternionic constant, which can depend on all the 
parameters of this two-particle wavefunction. Note that we have 
written this with independent time variables for the two particles.

These two-particle wavefunctions, with all their momentum-helicity 
labels, form a complete orthogonal set of functions in the space of 
$\textbf{x}_{1}$ and $\textbf{x}_{2}$. Note, however, that this product is 
ordered in a way that was meaningless in ordinary (complex) quantum 
theory but requires some extra bookkeeping in the quaternionic case.

Let's introduce some more compact notation for such wavefunctions.
\begin{equation}
\psi^{op}(1) \equiv exp(\eta_{1} \textbf{u}\cdot \hat{p_{1}}\textbf{p}_{1} 
\cdot\textbf{x}_{1})\; ||\; exp(\Omega_{1}t_{1})\label{g2}
\end{equation}
where the $||$ symbol separates those things that are to act on the left 
from what is to act on the right of whatever follows.
Then the two-particle wavefunction (\ref{g}) can be written simply as
\begin{equation}
\psi(1,2) = \psi^{op}(1)\;\psi^{op}(2)\;\phi;\label{g3}
\end{equation}
and we can also write the operator of the wave equation as
\begin{equation}
{\cal{D}} \equiv \frac{\partial}{\partial t} + 
\textbf{u}\cdot\nabla ||i - m || k.\label{g4}
\end{equation}

Next we have the propagators.
\begin{equation}
G(\textbf{x}-\textbf{x}', t-t') \equiv \sum_{p,\eta} exp(\eta 
\textbf{u}\cdot\hat{p}\textbf{p}\cdot (\textbf{x}-\textbf{x}')) ||
exp((i\eta p + km)(t-t'))\label{g5}
\end{equation}
where $\sum_{p,\eta} = \int_{H}\frac{d^{3}p}{(2\pi)^{3}}\sum_{\eta}$;
and this leads to
\begin{eqnarray}
&&{\cal{D}} G(\textbf{x}-\textbf{x}', t-t') = 0, \;\;\;\;\;
G(\textbf{x}-\textbf{x}', 0) = 
\delta^{3}(\textbf{x}-\textbf{x}')\label{g6} \\
&&G_{+}(x-x') \equiv \theta(t-t') 
G(\textbf{x}-\textbf{x}',t-t')\label{g7} \\
&&{\cal{D}}G_{+}(x-x') = \delta^{4}(x-x').\label{g8}
\end{eqnarray}
 The coordinate 
$x$ stands for the full space-time coordinates $t,\textbf{x}$. Now 
the equation (\ref{eb1}) can be briefly written as
\begin{equation}
\psi(x) = \int d^{4}x'\;G_{+}(x-x') s(x').\label{g9}
\end{equation}

Following that construction, we now  write down a general 
two-particle quaternionic wavefunction, as follows
\begin{equation}
\psi(x_{1},x_{2}) = \int 
d^{4}x_{1}'G_{+}(x_{1} - x_{1}')\; \int 
d^{4}x_{2}'G_{+}(x_{2} - x_{2}') s(x_{1}',x_{2}').\label{g10}
\end{equation}
Acting on this with two of those differential  operators gives
\begin{equation}
{\cal{D}}_{2}\;{\cal{D}}_{1}\;\psi(x_{1},x_{2}) = 
s(x_{1},x_{2}).\label{g11}
\end{equation}
This is a two-particle wave equation of the Bethe-Salpeter type, 
involving separate times as well as separate space coordinates.  The 
term $s$ might be left as an external source or might be used to 
represent some interaction, such as $V(1,2)\psi(1,2)$. Note that the 
order in which the two differential operators are applied is 
significant.

It seems easy now to extend this to any number of particles. This 
appears to be a significant advance over previous studies of 
quaternionic wave equations, although there are 
still  many issues to be faced.

\section {No Lagrangian}

If I use the interacting wave equation (\ref{e}), and think that 
$\psi^{*}$ is something independent of $\psi$, then 
 the following would be suggested as a Lagrangian density.
\begin{equation}
{\cal{L}} = i\psi^{*}\frac{\partial\psi}{\partial t}i - 
i\psi^{*}\textbf{u}\cdot\nabla\psi - 
i\psi^{*}e\varphi\psi + 
i\psi^{*}e\textbf{u}\cdot\textbf{A}\psi i - im\psi^{*}\psi 
e^{ieW}j.
\end{equation}
Varying $i\psi^{*}$ gives immediately the full wave equation 
for $\psi$. Before varying $\psi i$ on the right, we do a few things: 
partially integrate in space and time; move $i$ from left side to 
right side in the second and third terms and rearrange the $i$ and $j$ 
coefficients in the last term; this is justified because those 
$\psi^{*} \ldots \psi$ 
expressions are real. Then we get the adjoint wave equation.

But that prescription is not what the usual action principle allows. 
The familiar game from complex qm does not 
work here. If one varies each of the four real functions which make 
up both quaternionic functions $\psi$ and $\psi^{*}$, then we actually 
get 12 equations from the action principle. This is due to the fact 
that this Lagrangian is imaginary, that is, it consists of three 
imaginary parts and each of those parts must vanish after the variation.
If we write 
\begin{equation}
{\cal{L}} = i {\cal{L}}_{1} + j {\cal{L}}_{2} + k {\cal{L}}_{3},
\end{equation}
we find that the first term, ${\cal{L}}_{1}$, is Lorentz invariant 
(see equation (\ref{B9b})); but what should we do with the other two 
terms?

Our difficulty with a Lagrangian is different from the difficulty 
noted earlier for the Dirac-Majorana equation or for the pseudo-Weyl 
equation.  But we do have a problem here.

\section{Discussion}

Several advances have been made in trying to develop a sensible 
quantum theory based upon quaternions, rather than complex numbers.  So 
far, this work has been limited to the wave equation formalism. 

We have noted the lack of a conserved energy-momentum 
tensor (see Appendix A) as well as the lack of a Lagrangian. Nevertheless, we
 can write down the time-development operator as
\begin{equation}
U(t) = e^{Ht}, \;\;\;\;\; H = -\textbf{u}\cdot \nabla||i + 
m||k.\label{h}
\end{equation}
This operator $H$ commutes with the angular momentum operator 
$\textbf{J}$, but whether we want to call it the Hamiltonian is unclear.
Perhaps these questions wait for a full model of 
how this quaternionic wave system interacts with other physical 
systems.

Another approach that may be relevant to that problem, as well as to 
improving our treatment of many-particle systems, is the method of 
second quantization.  We are led to write down a quaternionic quantum 
field operator as
\begin{equation}
\psi(\textbf{x},t) = \sum_{p,\eta}\; exp(\eta 
\textbf{u}\cdot\hat{p}\;\textbf{p}\cdot\textbf{x})\;a_{p,\eta}\;
exp((i\eta p + km)t)\label{h1}
\end{equation}
involving some kind of annihilation/creation operators $a_{p, 
\eta}$.  With this we immediately get
\begin{eqnarray}
N = \frac{1}{(2\pi)^{3}}\int d^{3}x \;\rho = \sum_{p,\eta}\;
a^{\dagger}_{p,\eta}a_{p,\eta} \\
h = \frac{1}{(2\pi)^{3}}\int d^{3}x \;\tau = -\sum_{p,\eta}\;
a^{\dagger}_{p,\eta}a_{p,\eta}\;\eta p .
\end{eqnarray}
Can one be sure that the matrix product $a^{\dagger}a$ is real? If 
these are matrices in a Fock space of the sort we are familiar with, 
with nonzero elements only on one line parallel to the central 
diagonal, then this product is real. 

It remains unclear to this author whether the equation studied in this 
paper is merely an alternative mathematical formulation of things 
already well known or whether it may have consequential 
applications to some as-yet unidentified physics.

\vskip 0.5cm
\begin{center} \textbf{ACKNOWLEDGMENTS} \end{center}

 I am grateful to S. Adler and to K. Bardakci for helpful conversations.

 \vskip 0.5cm
\setcounter{equation}{0}
\def\theequation{A.\arabic{equation}}
\boldmath
\noindent{\bf Appendix A: General Tensors}
\unboldmath
\vskip 0.5cm

We can construct Lorentz covariant tensors of any rank, as follows.
Start with the direct product of the ``two-way'' derivative operators:
\begin{equation}
d^{n}_{\mu} = d_{\mu_{1}}\;d_{\mu_{2}} \ldots d_{\mu_{n}}\label{B}
\end{equation}
where the subscript $\mu$ now stands for the set of indices $\mu_{1} 
\ldots\mu_{n}$.
This expression is manifestly a covariant tensor of rank n as far as 
the coordinate transformations are concerned; and our task is to 
package these between the wavefunctions, which transform under 
infinitesimal Lorentz transformations as given in (\ref{c1}).

 We  note that 
the packages $\psi^{*}d^{n}_{\mu}\psi$ are real for n even and 
imaginary for n odd; while these conditions are reversed when we add 
the quaternions $\textbf{u}$ inside the package.

We find the following constructions for covariant tensors of rank n+1,
$Q^{(n+1)}_{\mu,\nu}=Q_{\mu,\nu} = (Q_{\mu,0}, \textbf{Q}_{\mu})$:
\begin{eqnarray}
Q_{\mu,\nu} = (\psi^{*}d^{n}_{\mu}\psi, 
\{\frac{-i}{2},\psi^{*}d^{n}_{\mu} \textbf{u}\psi\}) 
\;\;for\;\;n\;\;even \label{B1}\\
Q_{\mu,\nu} = (\{\frac{-i}{2},\psi^{*}d^{n}_{\mu}\psi\}, 
-\psi^{*}d^{n}_{\mu} \textbf{u}\psi) 
\;\;for\;\;n\;\;odd.\label{B2}
\end{eqnarray}
In the case $n=0$ this is just the 4-vector current, previously 
written as $j_{\mu}=(\rho,\textbf{U}_{1})$. All these tensors are real.

In addition, for n odd, we have the tensors of rank n,
\begin{equation}
R_{2,\mu} = \{\frac{-j}{2}, \psi^{*}d^{n}_{\mu}\psi\}, \;\;\;\;\;
R_{3,\mu} = \{\frac{-k}{2}, \psi^{*}d^{n}_{\mu}\psi\},\label{B3}
\end{equation}
which generalize the previously noted 4-vectors $D_{2}, D_{3}$.
And for n even, we have the tensors of rank n+2, $S_{\mu,\nu,\lambda} 
=- S_{\mu,\lambda.\nu}$:
\begin{equation}
S_{\mu,0,\alpha} = -S_{\mu,\alpha,0} =
\{\frac{-j}{2}, 
\psi^{*}d^{n}_{\mu}\textbf{u}_{\alpha}\psi\},\;\;\;\;\;
S_{\mu,\alpha,\beta}= \epsilon_{\alpha,\beta,\gamma} \{\frac{-k}{2},
 \psi^{*}d^{n}_{\mu}\textbf{u}_{\gamma}\psi\}\label{B4}
\end{equation}
where $\alpha, \beta,\gamma = 1,2,3$. This generalizes the 
previously noted six-vector $(\textbf{U}_{2},\textbf{U}_{3})$.

We can make lower rank tensors by contracting indices: 
\begin{equation}
g^{\mu_{1},\mu_{2}}Q^{(n+1)}_{\mu,\nu} = -(m^{2}+\frac{1}{4} 
\partial^{\lambda}\partial_{\lambda})
Q^{(n-1)}_{\mu',\nu}\label{B7}
\end{equation}
where $\mu_{1}$ and $\mu_{2}$ are in the set $\mu$; and the set 
$\mu'$ has these two indices removed. An alternative is to contract 
one of the $\mu$ indices with the $\nu$ index. We find, for solutions 
of the free wave equation (\ref{a}):
\begin{eqnarray}
&&g^{\mu_{1},\nu} Q^{(n+1)}_{\mu,\nu} = -m R^{(n-1)}_{3,\mu'} 
\;\; for \;\;n\;\;even \label{B8}\\
&&g^{\mu_{1},\nu} Q^{(n+1)}_{\mu,\nu} = 0 \;\; for 
\;\;n\;\;odd.\label{B9}
\end{eqnarray}

If we are looking to find a Lorentz scalar, a tensor of rank zero, take 
a closer look at the second rank tensor $Q_{\mu,\nu}$. The 
contraction is 
\begin{equation}
Q_{\mu}^{\mu} = \{\frac{-i}{2},\psi^{*}d_{0}\psi\} + 
\psi^{*}\textbf{u}\cdot \textbf{d}\psi = D_{1,0} + \tau \label{B9b}
\end{equation}
and this is exactly zero for the free equation (\ref{a}) but not for 
the extended equation (\ref{e}), where it equals $-ej_{\mu}A^{\mu}$.

Now we look at the contraction of such tensors with the derivative 
operator. In what follows we shall limit ourselves to solutions of the 
free equation (\ref{a}).  It is transparent that 
$\partial^{\mu_{1}}\;Q_{\mu,\nu} = 0$ for any $\mu_{1}$ in the set of 
labels $\mu$. The same holds true for the tensors R and S. 
Furthermore, by using the wave equation, one can show 
that 
\begin{eqnarray}
&&\partial^{\nu}\;Q_{\mu,\nu} =0\;\;for\;\;n\;\;even\label{B5} \\
&&\partial^{\nu}\;Q_{\mu,\nu} = 2m R^{(n)}_{2,\mu} \;\; 
for\;\;n\;\;odd.\label{B6}
\end{eqnarray}

For tensors of rank one, we have just the previously identified 
$j_{\mu}$, $D_{2,\mu}$,and  $D_{3,\mu}$, all of which are conserved.

At rank two, the usual desire is 
for a conserved symmetric tensor, which one can call the 
energy-momentum tensor. The closest we come here is the $Q_{\mu,\nu}$,
which is not symmetric and is conserved only on the first index. 
Nevertheless, this does allow us to write integral quantities which are 
conserved (their time derivatives vanish), as follows.
\begin{eqnarray}
&&V_{\nu} = (V_{0},\textbf{V}) \equiv \int d^{3}x\; 
Q^{(2)}_{0,\nu}\label{B10} \\
&&V_{0}  = -\int d^{3}x \; \psi^{*}\textbf{u}\cdot 
\textbf{d}\psi \label{B11}\\
&&\textbf{V} =\int d^{3}x\;(\textbf{D}_{1}+m\textbf{U}_{3}
- \frac{1}{2}\nabla \times \textbf{U}_{1})\label{B12}
\end{eqnarray}
This is not what we would identify as the energy-momentum, 
as noted at the end of Section 9.

For the second rank antisymmetric tensor we have
\begin{eqnarray}
&&\partial^{\mu}\;S_{\mu,\nu} = -2mj_{\nu} + 2 D_{3,\nu}\label{B13} \\
&&\tilde{S}_{\mu,\nu} = 
\epsilon_{\mu,\nu,\kappa,\lambda}\;S^{\kappa,\lambda}
/2\label{B14} \\
&&\partial^{\mu}\;\tilde{S}_{\mu,\nu} = -2 D_{2,\nu}\label{B15}.
\end{eqnarray}

\end{document}